\documentstyle[aps,pre,manuscript]{revtex}

\begin{document}

\draft

\title{Upper bound for the time derivative of entropy for nonequilibrium
stochastic processes}

\author{Bidhan Chandra Bag{\footnote{e-mail: pcbcb@yahoo.com}}
}

\address{Department of Chemistry, Visva-Bharati, Santiniketan 731 235, India}

\date{\today}

\maketitle

\begin{abstract}
We have shown how the intrinsic properties of a noise process can set an 
upper bound for the time derivative of entropy in a nonequilibrium 
system. The interplay of
dissipation and the properties of noise processes driving the 
dynamical systems in presence and absence of external forcing,
reveals some interesting extremal nature of the upper bound.

\end{abstract}



\section{Introduction}

A consequence of the second law of thermodynamics is that the rate of change
of entropy with time for a nonequilibrium stochastic process is always positive.
While in traditional classical thermodynamics, the specific nature of a
stochastic process is irrelevant, this may play an important role in 
understanding the connection between the phase space of a dynamical system 
and the related thermodynamically inspired quantities like entropy production, flux etc.
The relationship has recently been explored by a number of authors 
\cite{hol,leb,ruelle,jou,gaspard,bro,nik,fri,bb3,nicolis,bb1,bb2,sc1}. The
object of the present paper is to address a related issue.

In what follows we shall be concerned with the dissipative dynamical systems 
which are thermodynamically open \cite{lw}
in the sense that they can be described
by classical stochastic processes with the help of the standard Langevin
equations. When a dynamical system is driven by a noise process, e. g, color \cite{hr} or 
cross-correlated processes \cite{jz,jz3}, the nature of the noise processes may influence
the dynamical system through the appropriate modification of the phase space structure
of the overall system. In view of the immediate connection between 
information entropy and probability distribution function of the phase space 
variables, it is worthwhile to enquire about the imprints of the nature of noise
on entropy. Our specific aim in this paper is to show how the properties of the 
noise processes can set an upper bound on the rate of entropy change in a 
nonequlibrium system. By directly extending our earlier treatment on a
related problem \cite{bb1}
we have examined some interesting extremal properties of this
bound.

The outline of the paper is as follows: In Sec. II we introduce the
Fokker-Planck
description of a  dynamical system driven by two different
kinds of stochastic
processes (e. g, color and cross-correlated) and
an upper bound for the rate of entropy change 
based on this formulation. 
We illustrate the result in Section III for the two specific
cases. The paper is concluded in Section IV.

\section{The Fokker-Planck description of noise processes and upper bound for
the rate of entropy change}

We consider a dynamical system driven by the external Ornstein-Uhlenbeck
noise processes.
The relevant Langevin equations of motion can be written as

\begin{equation}
\dot{x_i} = F_i^0(\{x_i\})+\eta_i,  \; \; \; \; i=1, \cdots,N
\end{equation}

\noindent
where $N$ is the dimension of the phase space. $F_i^0(x)$ corresponds to 
the dissipative term as well as the external applied deterministic force, if any.
The second term $\eta_i$ in Eq.(1) refers to an external, Gaussian, color
noise for the i-th component of $x$. 

The Fokker-Planck equation corresponding to Langevin Eq.(1) 
in the extended phase space can be written as
[for details, see Ref.\cite{bb1}]
\begin{equation}
\frac{\partial \rho(X, t)}{\partial t} = - \sum_{i=1}^{2N} \frac{\partial}{\partial X_i}(F_i \rho)
+ \sum_{i=1}^{2N} D_i
\frac{\partial^2 \rho}{\partial X_i^2 } \; \; \;,
\label{e7}
\end{equation}

\noindent
where
\begin{eqnarray*}
X_i \left\{
\begin{array}{l}
= x_i \; \; \; {\rm for} \; i = 1, \cdots , N \\
= \eta_i \; \; \; {\rm for} \; i = N+1, \cdots , 2N \\
\end{array}\right.
\end{eqnarray*}

\noindent
$F_i$ and $D_i$ are drift and diffusion coefficients, respectively, and have
their usual significance as discussed in Ref. \cite{bb1}.
$\rho(X, t)$ is the extended phase space probability distribution function. 

As a second example we
consider a dynamical system driven by both additive and 
multiplicative noise processes $\eta_i$ and $\zeta_i$,
respectively. The Langevin equation for this process, in general, can be written as

\begin{equation} 
\dot{X_i} = L_i(\{X_i\}, t) + g_i(X_i) \zeta_i +\eta_i \; \; \; i=1, \cdots ,N
\end{equation} 

\noindent
$L_i$ contains the dissipative term as well as the external applied 
deterministic force, if any. $g_i(X_i)$ is the 
coupling between the system and the multiplicative processes $\zeta_i$.
$\zeta_i$ and $\eta_i$ are white, Gaussian
noise processes with the following correlation between them;

\begin{equation}
\langle \zeta_i(t) \eta_j(t') \rangle  = 
\langle \zeta_i(t') \eta_j(t) \rangle =
2 \lambda_{ij} \sqrt{D_{ij}'\alpha_{ij}} \delta (t-t') \delta_{ij}
\end{equation}

\noindent
$D_{ij}'$ and $\alpha_{ij}$ correspond to the strength of multiplicative 
and additive noises, respectively and $\lambda$ represents the cross correlation 
between them with the limit $0\le \lambda \le 1$. 

The Fokker-Planck equation corresponding to Langevin Eq.(3) can be written as

\begin{equation}
\frac{\partial \rho(X, t)}{\partial t} = - \sum_{i=1}^{N} \frac{\partial}{\partial X_i}(F_i \rho)
+ \sum_{i=1}^{N} D_i
\frac{\partial^2 \rho}{\partial X_i^2 } \; \; \;.
\label{e10}
\end{equation}

Again $F_i$ and $D_i$ are drift and diffusion coefficient, respectively,
and have same significance as in the Eqs. (11-13) of Ref.\cite{bb1}.

The Fokker-Planck Eqs.(2) or (5) can be rearranged into the 
general form of 
continuity equation
\begin{equation}
\frac{\partial \rho(X,t)}{\partial t} = -\nabla_X . j
\end{equation}

where $j$ denotes the current and $\nabla_X$ term the phase space divergence. The
$i$-th component of $j$ can be written as

\begin{equation}
j_i =F_i \rho -D_i \frac{\partial \rho}{\partial X_i}
\end{equation}

Using Eq.(6) we are now in a position to define
the upper bound for the rate of
evolution of entropy.
In the microscopic picture, the Shannon form of the entropy is connected 
to the probability density function $\rho(X,t)$ as

\begin{equation}
S = -\int dX \rho(X,t) \ln \rho(X,t) \; \;.
\label{e14}
\end{equation}

The time evolution equation for entropy can then be written as

\begin{equation}
\frac{dS}{dt} = \int dX \nabla_X.j \ln \rho \; \;,
\end{equation}

\noindent
where Eq.(6) is used.

Integrating Eq.(9) by parts, one obtains
\begin{equation}
\frac{dS}{dt} = -\int dX \frac{1}{\rho} j.\nabla_X \rho \; \;,
\end{equation}

\noindent
where we have used following boundary conditions \cite{bro}

\begin{equation}
j|_{boundary} = 0
\end{equation}

\noindent
and

\begin{equation}
j \ln\rho|_{boundary} = 0
\end{equation}

In the next step an application of the Schwartz inequality 
$|\int dX g h|^2 \leq \int dX |g|^2 \int dX |h|^2$ to the integral (10)
where $g$ and $h$ can be appropriately identified
yields an upper bound for the rate of entropy change

\begin{eqnarray*}
\frac{dS}{dt} \leq U_B(t)
\end{eqnarray*}

\begin{equation}
U_B(t) = \left(\int dX \frac{j^2}{\rho}\right)^{1/2} \times 
\left(\int dX \frac{(\nabla_X \rho)^2}{\rho}\right)^{1/2}
\end{equation}

It is important to note that the second integral is same as the trace of 
Fisher information matrix \cite{fri} and this inequality is valid if and only if 
$j / \sqrt{\rho}$ is not a constant multiple of 
$(\nabla_X \rho)/\sqrt{\rho}$. To find the explicit time dependence
of the upper bound we work out simple examples for each of the noise processes in the 
next section.

\section{Applications}

\subsection{The upper bound for 
a dynamical system driven by an external color noise}

As a simple illustration, we consider a Langevin equation of motion for
a dissipative dynamical system
driven by an external, Gaussian Ornstein-Uhlenbeck
noise, $\eta_1$. 

\begin{equation}
\dot{X_1} =  -\gamma X_1 + f_c+ \eta_1
\end{equation}

The noise correlation of $\eta_1$ is given by

\begin{equation}
\langle \eta_1 (t) \eta_1 (t') \rangle = \frac{D^0}{\tau} 
\exp \left ( -\frac{|t-t'|}{\tau} \right )
\end{equation}

$\gamma$ in Eq.(14) is the dissipative parameter and $f_c$ is a constant
external applied force term which is used to identify specific interplay between
$\gamma$ and $\tau$. 

For the Langevin Eq. (14) the Fokker-Planck Eq.(2) becomes (see Ref.[11])

\begin{eqnarray}
\frac{\partial \rho}{\partial t} & = & \gamma \frac{\partial X_1 \rho}{\partial X_1}
-f_c \frac{\partial \rho}{\partial X_1} -X_2 \frac{\partial \rho}{\partial X_1}
+\frac{1}{\tau} \frac{\partial X_2\rho}{\partial X_2} + \frac{D^0}{\tau^2}
\frac{\partial^2 \rho}{\partial X_2^2} 
\end{eqnarray}

\noindent
where $X_2=\eta_1$.

We now use the following transformation in the Eq.(16)
\begin{equation}
U=a X_1+ X_2  \; \; \;,
\end{equation}

where $a$ is a constant to be determined.

Then  Eq.(16) reduces into the following one dimensional form:
\begin{equation}
\frac{\partial \rho(U,t)}{\partial t} = \frac{\partial}{\partial U}(\Gamma U)\rho 
-F_u \frac{\partial \rho}{\partial U}+ D_s\frac{\partial^2 \rho}{\partial U^2}  \; \; ,
\end{equation}

\noindent
where 
\begin{equation}
\Gamma U = \gamma a X_1 -a X_2 + \frac{X_2}{\tau} \; \; , \; \;
F_u=af_c \; \; {\rm and} \; \;
D_s = \frac{D^0}{\tau^2} \; \; .
\end{equation}

Here $\Gamma$ is again a constant to be determined. Using Eq.(17) in
Eq. (19) and comparing the coefficients of $X_1$ and $X_2$  we find

\begin{equation}
\Gamma  = \gamma  \; \; {\rm and} \; \; 
a =  \frac{1-\gamma \tau}{\tau} \; \; .
\end{equation}

We then search for the Green's function or conditional probability
solution for the system at $U$ at time  $t$ for the initial condition
given by

\begin{equation}
\rho(U, t=0) = \lim_{\epsilon \rightarrow \infty} \frac{\epsilon}{\pi}
\exp[-\epsilon (U-U')^2]
\end{equation}

We now look for a solution of the Eq.(18) of the form

\begin{equation}
\rho(U, t|U', 0) = \exp[G(t)]
\end{equation}

\noindent
where
\begin{equation}
G(t) = -\frac{1}{\sigma(t)} (U-\beta(t))^2 +\ln \nu(t)
\end{equation}

We will see that by suitable choice of $\beta(t), \sigma(t), \nu(t)$ we can
solve Eq.(18) subject to the initial condition

\begin{equation}
\rho(U, 0|U', 0) = \lim_{\epsilon \rightarrow \infty} \frac{\epsilon}{\pi}
\exp[-\epsilon (U-U')^2] \; \;.
\end{equation}

Comparison of this with (22) and $G(0)$ shows that

\begin{equation}
\sigma(0)=\frac{1}{\epsilon}, \; \; \; \beta(0)=U', \; \; \;  \nu(0)=\frac{\epsilon}{\pi} \; \;.
\end{equation}

If we put (22) in (18) and equate the coefficients of equal powers of $U$
we obtain after some algebra the following set of equations

\begin{equation}
\dot{\sigma(t)} = -2\Gamma \sigma(t) +4D_s
\end{equation}

\begin{equation}
\dot{\beta(t)} =  -\beta(t) +F_u
\end{equation}

\begin{equation}
\frac{1}{\nu(t)}\dot{\nu(t)} = -\frac{1}{2\sigma(t)}\dot{\sigma(t)}
\end{equation}

The relevant solutions of $\sigma(t)$ and $\beta(t)$ for the present problem which
satisfy the initial conditions as stated earlier are given by

\begin{equation}
\sigma(t) = \frac{2D_s}{\Gamma}(1-\exp(-2\Gamma t))+ \sigma(0) \exp(-2\Gamma t)
\end{equation}

\noindent
and

\begin{equation}
\beta(t) = \frac{F_u}{\Gamma}(1-\exp(-2\Gamma t))+ \beta(0) \exp(-\Gamma t)
\end{equation}

Now making use of Eqs. (22), (29) and (30) in (13) we finally obtain explicit
time dependence of the upper bound $U_B(t)$ for the rate of entropy
change as

\begin{eqnarray*}
\frac{dS}{dt} \leq U_B(t)
\end{eqnarray*}

\noindent
where


\begin{equation}
U_B(t) =\frac{\left(2 \beta^2 \Gamma^2 \sigma -4\beta \Gamma F_u \sigma+
2F_u^2\sigma +\Gamma^2\sigma^2+4D_s^2 -4D_s\Gamma \sigma \right)^{1//2}}{\sigma}
\end{equation}


We now examine specifically the long time limit, i. e, $t \rightarrow \infty$
of the above result (31). As $t \rightarrow \infty$ Eqs. (29) and (30) 
reduce to

\begin{equation}
\sigma(\infty) = \frac{2D_s}{\Gamma} \; \; {\rm and} \; \;
\beta(\infty) = \frac{F_u}{\Gamma} \; \; .
\end{equation}

It is easy to check that as $t \rightarrow \infty$
the numerator of the right hand side of Eq.(31) vanishes
both in presence or absence of $F_u$.
Therefore we obtain the equation

\begin{equation}
\frac{dS}{dt} = 0
\end{equation}

This equality holds since in the long time limit
$j=0$ (see Eq. 18).
At any other time the time dependence of the 
upper bound $U_B$ for the rate of entropy change is explicitly
shown in Fig.1. We choose the initial conditions  $\sigma(0) =0, \beta(0) =1.0$
and parameter values $D^0 =1.0, f_c =1.0, \gamma=1.0$ and $\tau =1.0$.
Fig.1 shows that except for an initial short period 
$U_B(t)$ decreases almost exponentially with time. In absence of
$f_c$ the time dependence of $U_B$ follows a similar pattern. In Fig.2(a-b)
we plot $U_B$ at $t=5$ vs correlation time $\tau$ in absence and presence
of the external forcing $f_c$. As expected $U_B$ increases monotonically with 
the $\tau$ [in Fig.2(a)] which is a clear signature of the persistence of the
nonequilibrium situation in  contrast to the case in Fig.2(b) where the interplay
of $\tau$ with external forcing $f_c$ results in a minimum in
$U_B$. The result of Fig.2(b) is qualitatively same to that of the Fig.1 of Ref.[11] 
where only entropy production in the stationary state is considered. In 
the present context however the upper bound of the sum of entropy production
and entropy flux [11] at any arbitrary time is considered. The relation
between entropy flux ($E_F$) and entropy production ($E_P$) in the long time 
limit for the present model [11] is

\begin{equation}
E_P = - E_F \; \; \; = \frac{(1-\gamma \tau)^2 f_c^2}{D^0} \; \; .
\end{equation}

\noindent
Using  above equation in Eq.(31) at time $t\rightarrow \infty$ we have

\begin{equation}
U_B = [2\gamma E_P+ 2 \gamma E_F]^{1/2} \; \; = 0.
\end{equation}

Since near equilibrium $E_P$ approaches $-E_F$ the upper 
bound of time derivative of entropy as shown in Fig.2(b) 
mimics the result
of Fig.1 of Ref.[11].

In Fig. 3(a-b) we plot the variation of $U_B$ (at $t=5.0$)
vs dissipative constant $\gamma$ in absence (3a) and presence (3b) of the external
force $f_c$. While an increase in $\gamma$ facilitates the approach to
stationarity as evident from the monotonic  decrease of the bound in Fig. 3(a),
its effect becomes more interesting when the external $f_c$ is switched on
(Fig. 3(b)). One observes that the bound passes first through minimum
followed by a maximum to settle down at the vanishing level for the 
large values of dissipation. 
It is thus apparent that in presence of the external constraint the
properties of the noise processes and the dynamical characteristics
of the system play an important part for the upper bound for the rate of entropy
change.

\subsection{The upper bound in a cross-correlated noise driven system}

We now turn to the second case where a simple dissipative system is 
driven by both additive and multiplicative noises.

\begin{equation}
\dot{X_1} =  - \gamma X_1 -\zeta_1 X_1 +\eta_1
\end{equation}

\noindent
Eq.(5) for this system reduces to ( for details, see Ref.\cite{bb1} )
\begin{equation}
\frac{\partial \rho(X_1, t)}{\partial t} = -\frac{\partial (F_1 \rho)}{\partial X_1}
+ D_1 \frac{\partial^2 \rho}{\partial X_1^2}
\end{equation}

\noindent
where the drift term is

\begin{equation}
F_1 = -(\gamma + 2 D_{11}' -\nu) X_1 + l
\end{equation}

\noindent
with
\begin{equation}
l = (2-\nu) \lambda_{11} \sqrt{D_{11}' \alpha_{11}} \; \; ,
\end{equation}

\noindent
and


\begin{equation}
D_1 = \frac{\left[\alpha_{11} \gamma^2 + (2-\nu) D_{11}' \alpha_{11} \{(2-\nu)D_{11}'
+2\gamma -2 \gamma \lambda_{11}^2 -\lambda_{11}^2 (2-\nu) D_{11}' \} \right]}
{\Gamma'^2}
\end{equation}


\noindent
where

\begin{equation}
\Gamma' = \gamma + 2 D_{11}' -\nu \; \; .
\end{equation}

\noindent
Here $D_{11}^\prime$ and $\alpha_{11}$ are the multiplicative and additive
noise strength respectively. $\lambda_{11}$ is the cross-correlation between
the noise processes. $\nu=1$ stands for the Startonovich and $\nu = 0$ for the
Ito convention, respectively.

The Eq.(37) is very much similar to Eq. (18). Hence, the upper bound for the rate
of entropy change can be calculated as in the previous case and the final expression 
for the upper bound $U_B$ is given by

\begin{eqnarray*}
\frac{dS}{dt} \leq U_B(t)
\end{eqnarray*}

\noindent
where


\begin{equation}
U_B(t) =\frac{\left(2 \beta_1^2 \Gamma'^2 \sigma_1 -4\beta_1 \Gamma' l \sigma_1+
2l^2\sigma_1 +\Gamma'^2\sigma_1^2+4D_1^2 -4D_1\Gamma' \sigma_1 \right)^{1//2}}{\sigma_1}
\; \; .
\end{equation}


\noindent
Here the time evolution of $\sigma_1(t)$ and $\beta_1(t)$ can be written as

\begin{equation}
\sigma_1(t) = \frac{2D_1}{\Gamma'}(1-\exp(-2\Gamma' t))+ \sigma_1(0) \exp(-2\Gamma' t)
\end{equation}

\noindent
and

\begin{equation}
\beta_1(t) = \frac{l}{\Gamma'}(1-\exp(-2\Gamma' t))+ \beta_1(0) \exp(-\Gamma' t)
\end{equation}

The initial conditions for $\sigma_1(0)$ and $\beta_1(0)$ can be chosen as 
in Eq.(25). $l, D_1$ and $\Gamma'$  are determined by Eqs. (39), (40) and (41).
Again it is easy to check that for the correlated noise process under stationary
condition we obtain the usual equality

\begin{equation}
\frac{dS}{dt} = 0 \; \;.
\end{equation}

The time dependence of $U_B$ for a correlated noise process (we fix the
parameter values as $\gamma =1.0, D_{11} =1.0, \lambda_{11} = 0.5,$ 
$\alpha_{11} =1.0$ and the initial conditions $\sigma_1(0)=0, \beta_1(0)=0$) is more or
less same as that of Fig.1. In Fig.4 we exhibit the variation of $U_B$
(at $t=5.0$) with the strength of correlation $\lambda_{11}$. It is interesting
to note that although both multiplicative and additive noises are
independently and instantaneously correlated their mutual strength
of correlation $\lambda_{11}$ drives the system away from stationarity
more strongly (as compared to the case  corresponding to the 
variation of correlation time $\tau$ in Fig. 2(b)). 
No minimum, however, is obtained. We mention, in passing, that since the models
considered here are linear and are exactly solvable by Green's function
of Gaussian form the computed upper bound is an exact one.

\section{Conclusions}
Based on Fokker-Planck description of color and cross-correlated noise-driven dynamical
systems we have shown how the intrinsic properties of a noise process can set an
upper bound for the rate of entropy change in a nonequilibrium system. Since
the dissipative forces tend to equilibrate the system while an increase in
the noise correlation time in a color noise process or an increase in the
strength of correlation in
cross-correlated noise processes acts in the opposite direction, an interplay of
them makes the dynamical system exhibit interesting extremum properties of this 
upper bound. This is manifested in the maxima and minima of the bound for the time
derivative of Shanon entropy as a function of the strength of dissipation,
correlation time or strength of correlation in presence or absence of the external
forces acting on the dynamics. Since the color and cross-correlated
noise processes occur in many situations in physics and chemistry, the
observation made in this paper, we hope, will be useful for  understanding 
the close connection between irreversible thermodynamics and dynamical systems in many
related issues.

\acknowledgments
The author expresses his deep sense of gratitude to D S Ray for suggesting 
the present problem and for his kind attention throughout its progress.

\newpage
\begin{center}
{\bf Figure captions}
\end{center}

\noindent
{\bf Fig.1.}
Plot of upper bound for the time derivative of entropy  
$\ln U_B(t)$ vs time $t$ 
for the Eq.(31) using 
$\gamma =1.0$, $f_c = 1.0$, $D^0 = 1$, $\tau=1.0$, 
$\beta(0)=1.0$ and $\sigma(0)=0.0$ (Units are arbitrary).

\noindent
{\bf Fig.2(a).}
Plot of  $U_B$ vs correlation time $\tau$ at $t=5.0$
for the Eq.(31) using $f_c = 0.0$ and values of other parameters 
same as in Fig.1(Units are arbitrary).

\noindent
{\bf Fig.2(b).}
Same as in Fig.2(a) but for $f_c=1.0$ (Units are arbitrary).

\noindent
{\bf Fig.3(a).}
Plot of  $U_B$ vs dissipative constant $\gamma$ at $t=5.0$
for the Eq.(31) using $\tau=1.0$ and values of other parameters 
same as in Fig.2(a) (Units are arbitrary).

\noindent
{\bf Fig.3(b).}
Same as in Fig.3(a) but for $f_c=1.0$ (Units are arbitrary).

\noindent
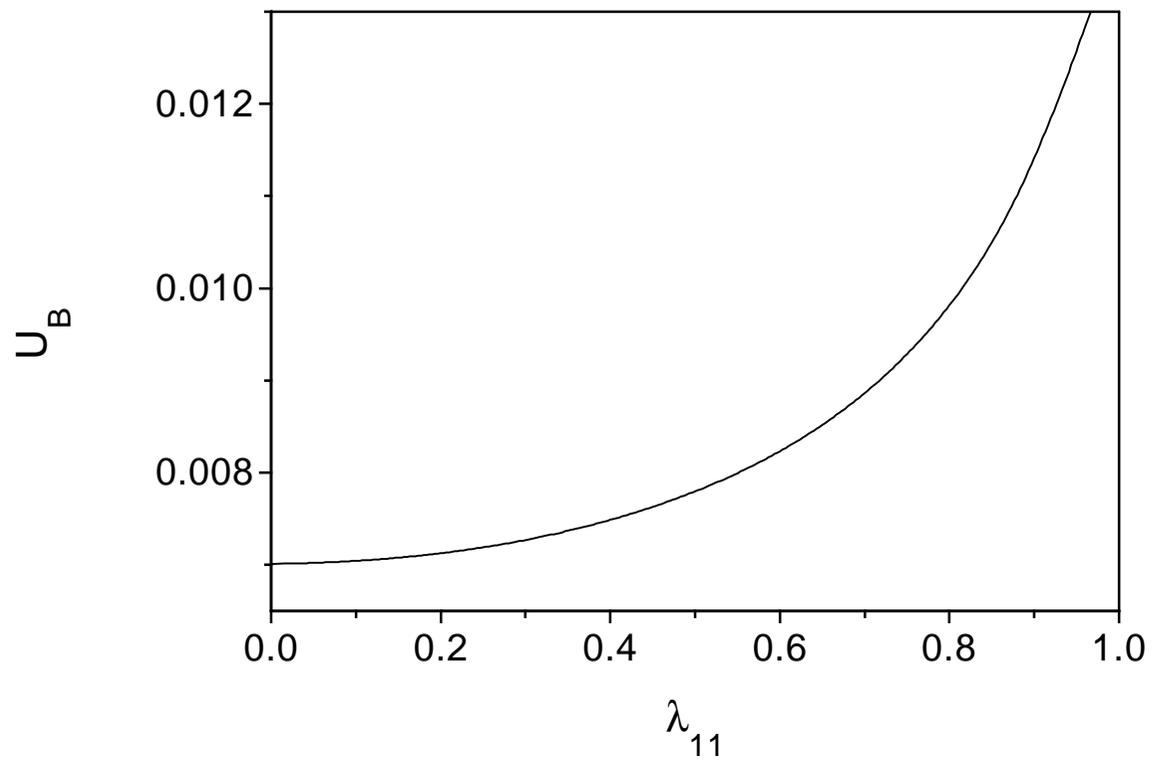
{\bf Fig.4.}
Plot of  $U_B$ vs noise correlation strength $\lambda_{11}$ at $t=5.0$
for the Eq.(42) using 
$\gamma =1.0$, $D_{11} = 1$, $\alpha_{11}=1.0$, 
$\beta_1(0)=1.0$ and $\sigma_1(0)=0.0$ (Units are arbitrary).


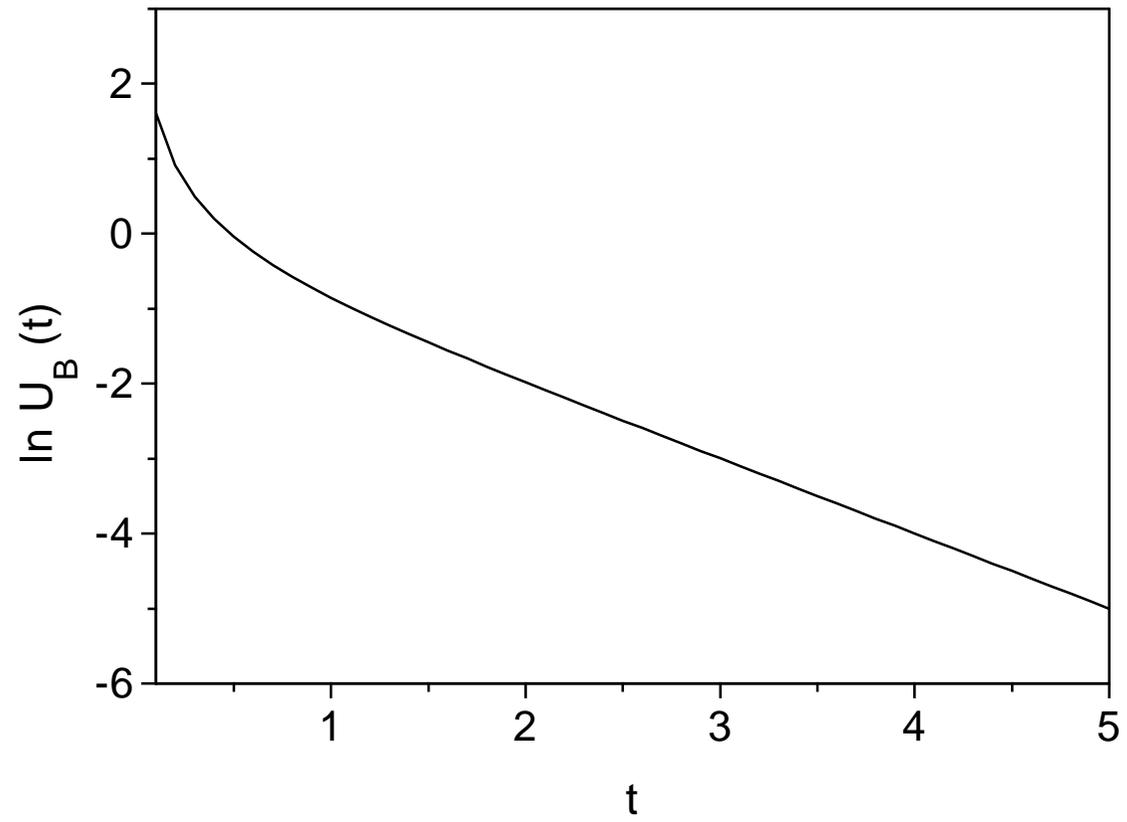
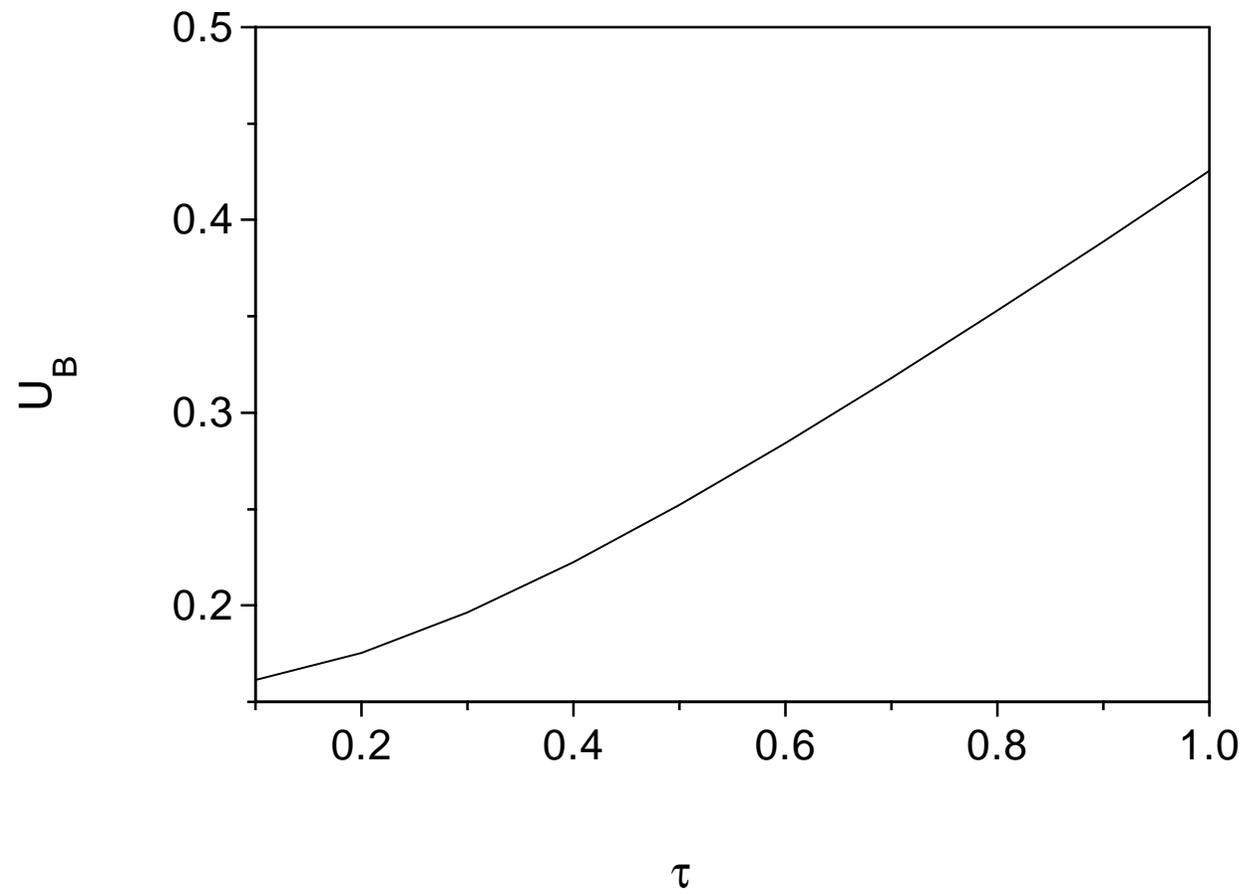
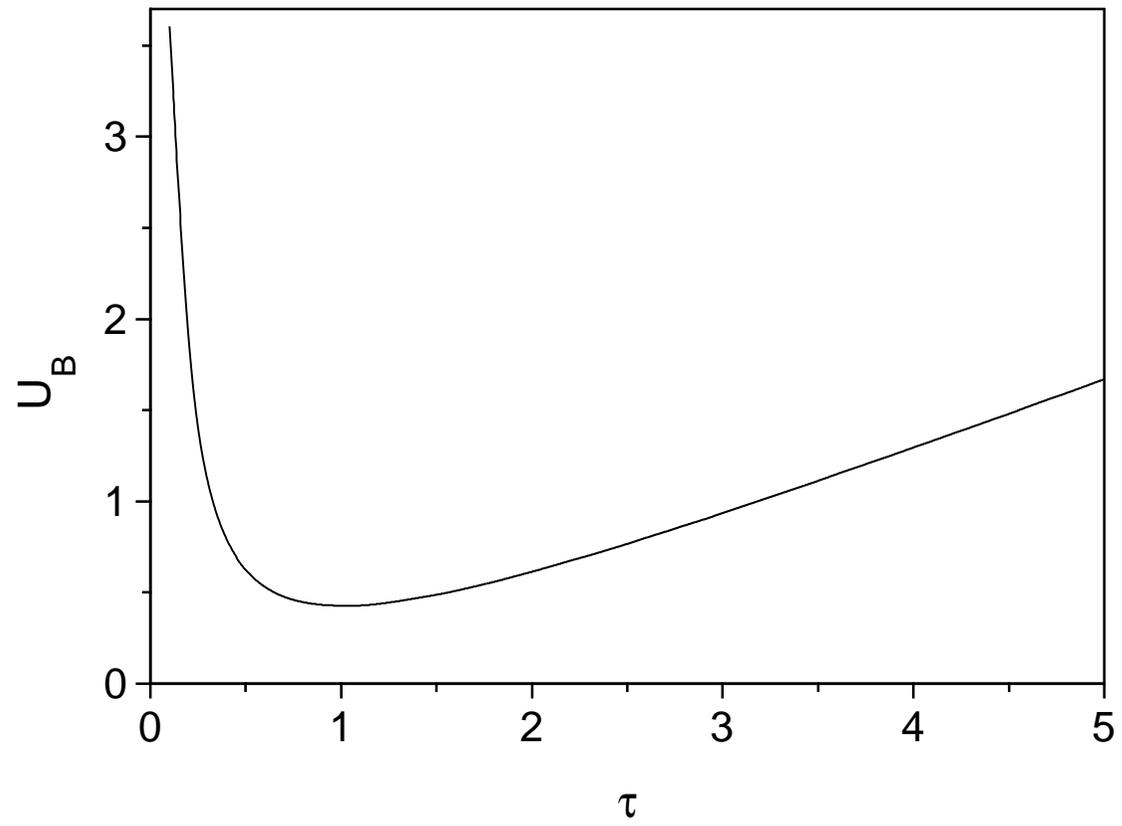
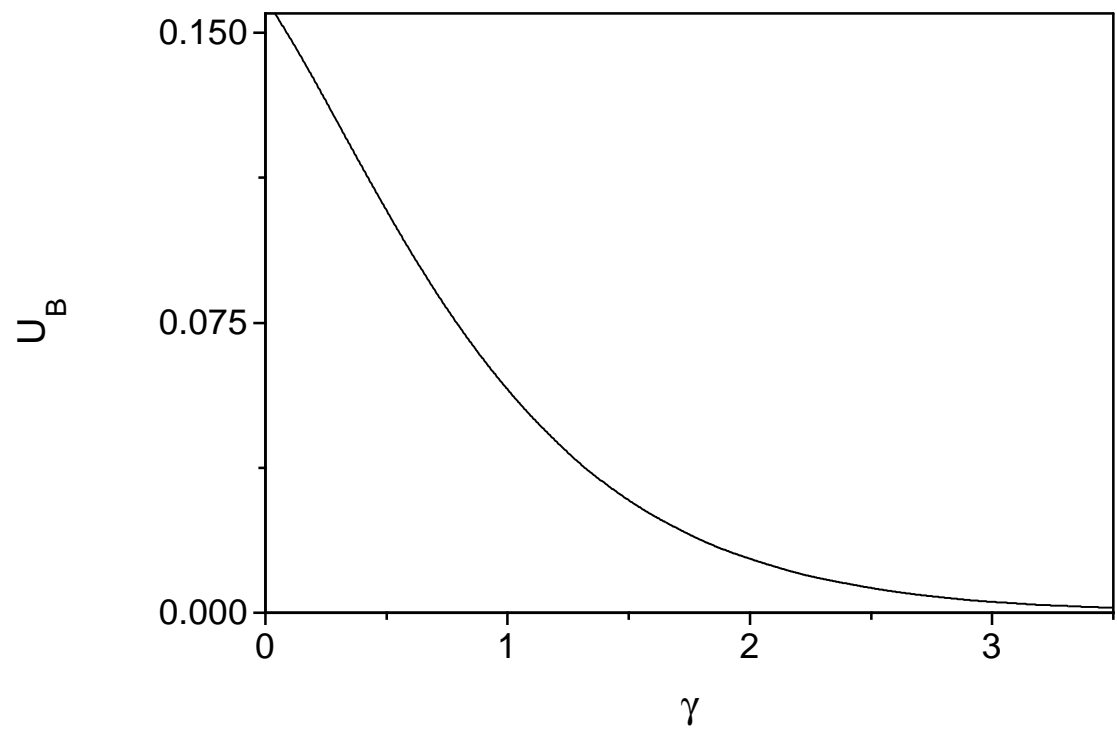
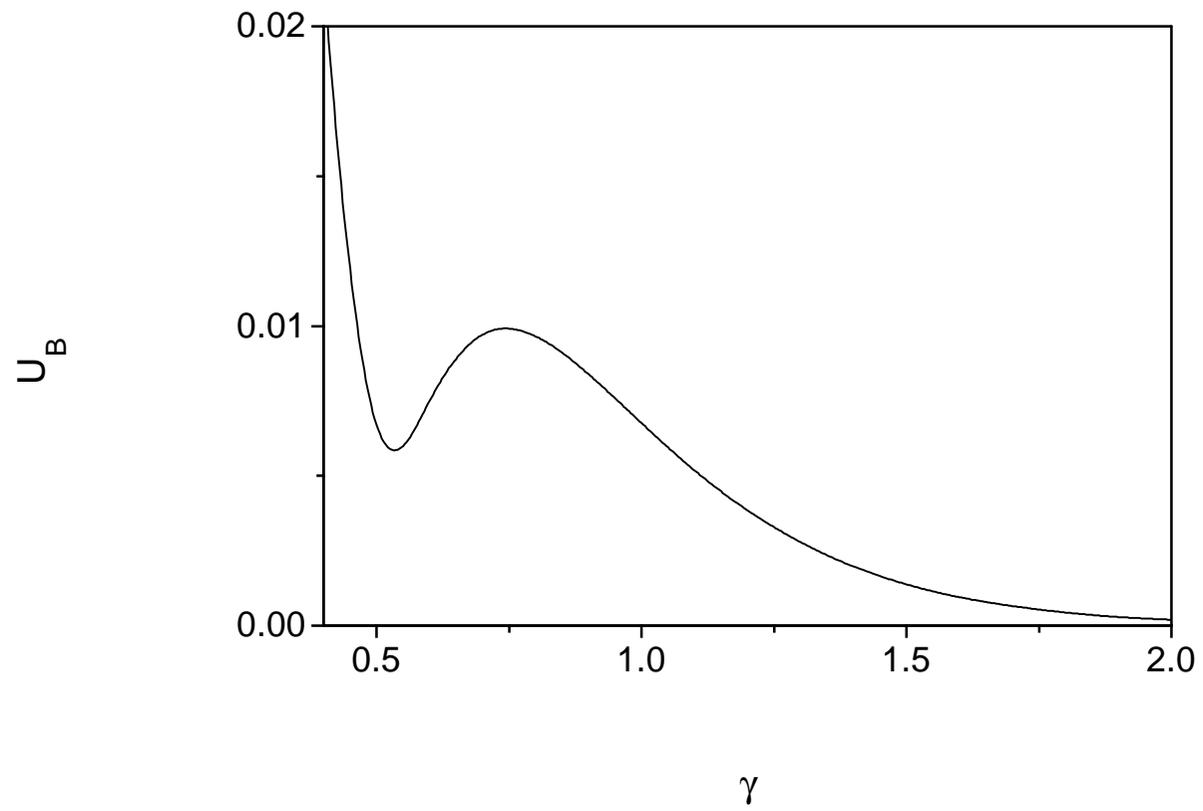
\end{document}